\documentclass[aps,prd,showpacs,floats,nofootinbib,superscriptaddress,preprint,tightenlines]{revtex4}

\usepackage{amsmath}
\usepackage{amsfonts}
\usepackage{amssymb}
\usepackage{pictex}
\usepackage{epsfig}

\usepackage[latin1]{inputenc}

\def\be{\nopagebreak[3]\begin{equation}}
\def\ee{\end{equation}}
\def\ba{\nopagebreak[3]\begin{eqnarray}}
\def\ea{\end{eqnarray}}
\def\bas{\nopagebreak[3]\begin{eqnarray*}}
\def\eas{\end{eqnarray*}}

\def\d{{\rm d}}

\def\a{\alpha}
\newcommand{\teta}{\rlap{\lower2ex\hbox{$\,\tilde{}$}}\eta{}}

\newcommand{\bi}{\begin{itemize}}
\newcommand{\ei}{\end{itemize}}

\def\l{\lambda}

\def\g{\gamma}
\def\k{\kappa}

\usepackage{enumerate}

\def\b{\beta}

\def\a{\alpha}

\def\t{\tau}
\def\k{\kappa}

\begin{document}

\title{On the choice of time in the continuum limit of polymeric effective theories} 

\author{Alejandro Corichi}\email{corichi@matmor.unam.mx}
\affiliation{Centro de Ciencias Matem\'aticas,
Universidad Nacional Aut\'onoma de
M\'exico, UNAM-Campus Morelia, A. Postal 61-3, Morelia, Michoac\'an 58090,
Mexico}
\affiliation{Center for Fundamental Theory, Institute for Gravitation \& the Cosmos,
Pennsylvania State University, University Park
PA 16802, USA}
\author{Tatjana Vuka\v sinac}
\email{tatjana@umich.mx}
 \affiliation{Facultad de
Ingenier\'\i a Civil, Universidad Michoacana de San Nicolas de
Hidalgo\\ Morelia, Michoac\'an 58000, Mexico}

\begin{abstract} 
In polymeric quantum theories, a natural question pertains to the so called continuum limit, corresponding to the 
limit where the `discreteness parameter' $\lambda$ approaches zero. In particular one might ask whether the limit 
exists and, in that case, what the limiting theory is. Here we review recent results on the classical formulation 
of the problem for a soluble model in loop quantum cosmology. We show that it is only through the introduction of 
a particular $\l$-dependent internal time function that the limit $\l\to 0$ can be well defined. We then compare 
this result with the existing analysis in the quantum theory, where the dynamics was cast in terms of an internal 
($\l$-independent) parameter for which the limit does not exist. We briefly comment on the steps needed to define 
the corresponding time parameter in the quantum theory for which the limit was shown to exist classically. 
\end{abstract}

\maketitle

\section{Introduction}

In the study of simple cosmological models from the perspective of loop quantum gravity, a non-standard, {\it polymeric} 
quantum theory \cite{exotic} has proven to be essential 
\cite{lqc}. One of the main features of this class of quantum theories is that they depend on a new dimensionful parameter 
$\l$. In the case of simple quantum mechanical systems 
like the harmonic oscillator \cite{AFW,CVZ2}, this parameter has information about the discretization of space; 
the position operator becomes discrete in equidistant steps given by $\l$. 
In loop quantum cosmology (LQC) the parameter has information about the `discreteness' of the loop geometry encoded in 
the fact that one considers finite holonomies to approximate the curvature of the connection. Since the standard Wheeler 
DeWitt (WDW) quantum theory can be recovered by setting $\l=0$ in the quantum constraint, one might wonder what is the corresponding 
limit of dynamical physical states. Do we recover the WDW states in the limit? 

In the case of a simple model in loop quantum cosmology, a spatially flat FRW model coupled to a massless scalar field, 
there has been a certain amount of results in this direction. In \cite{slqc} the quantum model was shown to be exactly soluble 
and the issue of the limit $\lambda\to 0$ was addressed. In that case, the choice of internal time was rather natural since the 
physical quantum states could be interpreted as {\it 
evolving} with respect to an internal time given by the massless scalar field $\phi$. It was shown there that the 
limit $\l\to 0$ was not uniform, {\it with respect to the internal time $\phi$}, so the expectation value of the volume, 
as function of $\phi$, does not approximate the corresponding value of the WDW theory for all times. One does not recover 
the WDW  theory in this limit \cite{slqc,CVZ3}. The question remained open as to whether the limit exists for a different 
choice of internal time.

A simpler and conceptually less sophisticated question pertains to the corresponding limit in the 
classical domain. There, a one parameter family of classical constrained systems, labeled by $\l$, exists. They contain 
the dynamical information of semiclassical states of the quantum theory, and have sometimes been referred to as the 
{\it effective theory}. In other instances one could simply consider a family of such classical theories, without reference 
to a particular quantum theory. It is then natural to ask whether the limiting process $\l\to 0$ is well 
defined for these theories. Borrowing the nomenclature from the quantum theory, we refer to this limit as the 
{\it continuum limit} of the effective theories (see \cite{CV} for further discussion on this issue).

Recently, this problem involving classical theories was studied in detail \cite{CV}. It was argued that, in order to have 
a well defined limit, one needs to introduce an {\it internal time function} as a physical Dirac observable with respect 
to which the dynamics can be described. Furthermore, the corresponding time function has to depend explicitly on the 
parameter $\l$ for the limit to be well defined, and the dependence is severely limited. In particular, it was shown that, 
for the LQC example, the internal time given by the scalar field is not a good parameter for this purposes. That is, the 
limit is not well defined in that case, in complete analogy with the quantum result of \cite{slqc}.

The purpose of this contribution is to make the relation between these two results transparent. In the first part we review 
the classical theory and explicitly construct the 
preferred time function that, interestingly, turns out to be the parameter that, in the spacetime interpretation, measures 
proper time of co-moving observers. We show that with 
respect to proper time, the limit is well defined. In the second part we briefly review the quantum theory and explicitly 
show the $\lambda$-dependence of the expectation value of 
the volume operator as a function of the scalar field $\phi$. 
We compare the expectation values on different scales and give a map between the corresponding states for which the expectation 
values are close to each other for large positive values of the scalar field, but then they grow apart
as $\phi$ gets smaller. A different map is needed to 'synchronize' the expectation values for large negative $\phi$, but as
$\phi$ goes away from this asymptotic region the expectation values get further apart. We can not simultaneously synchronize the
both branches, because the bounce occurs for different values of $\phi$ at each scale.
We also show that, taking the limit $\l\to 0$, one can indeed recover either branch of the 
WDW theory (contracting or expanding) from the LQC states and 
approximate the dynamics (as described by expectation values) far from the bounce. But, as noted earlier 
\cite{slqc,CVZ3}, one can not recover the whole dynamical trajectory since 
the LQC dynamics has always a bounce, while the WDW theory follows the classical trajectory into the singularity.\footnote{This has to be contrasted with some simple unconstrained quantum mechanical systems where the limit has been shown 
to exist \cite{AFW,CVZ2,CVZ1,jorma}.}
We end with some comments on the open problem of how to define proper time in the quantum theory with respect to which 
the convergence question can be posed.

\section{Classical theory}

We are interested in completely constrained system with one constraint $\cal C$, that
generates the time evolution, which is actually a gauge transformation.   
We denote by $\cal M$ the $2n$ dimensional phase space of this system, with canonical coordinates $\{ q^a,p_a\}$, 
$a=1,\dots n$. 
For this system we need to solve the constraint ${\cal C}(q^a,p_a)\approx 0$ and to find gauge invariant phase space functions,
i.e. Dirac observables. We shall follow the ideas of Rovelli and Dittrich, who associate to every
pair of phase space functions a one-parameter family of Dirac observables \cite{rovelli,dittrich}. 
This construction starts with the definition of the flow $\alpha^t_C$ generated by the constraint, where $t$ 
is an evolution parameter associated to ${\cal C}$.
For an arbitrary smooth phase space function $F(q^a,p_a)$ it can be calculated as

\be
\alpha^t_C(F)=\sum_{n=0}^{\infty}\frac{t^n}{n!}\{ {\cal C},F\}_n\, ,
\ee
where $\{ {\cal C},F\}_0=F$ and $\{ {\cal C},F\}_{n+1}=\{ {\cal C},\{ {\cal C},F\}_n\}$.

A partial observable is an arbitrary phase space function, to which in principle one can associate a number after
some measuring procedure. It does not have to be a Dirac observable.
We first choose as a partial observable the phase space function $R(q_a,p_a)$ 
that can be used to measure the "time" along the gauge orbits, generated by the constraint.
In order to be a good "clock" variable the function $R(t)\equiv \alpha^t_C(R )$ has to be invertible.
We shall consider another phase space function $F(q^a,p_a)$ and calculate its value when $R(t)=t_0$

\be
F\vert_{t_0}\equiv \alpha^t_C(F)\vert_{R(t)=t_0}\, .
\ee
Since $R(t)$ is an invertible function, $F\vert_{t_0}$ form a one parameter family
of (complete) Dirac observables, i.e. $\{ F\vert_{t_0},{\cal C}\} =0$ (For details, see \cite{dittrich}). It is obvious that
we can choose different "clock" variables, and the resulting family of Dirac observables, in general, are not
equivalent. We are going to exploit this idea and analyze two different choices for $R(q_a,p_a)$ and the
corresponding Dirac observables and we shall argue that one of them is more convenient in order to study the
convergence of the suitable constructed effective theories.
 
As our first example of a totally constrained classical system we shall consider 
a $k=0$ FRW cosmological model coupled to a massless scalar field, and we shall construct one family
of Dirac observables. In this case the metric takes the form

\be
ds^2=-N(t)^2dt^2+a(t)^2\delta_{ij}dx^idx^j\, ,
\ee
where $N(t)$ is (arbitrary) lapse function and $a(t)$ is the scale factor.  
The classical phase space for the gravitational sector can be described by two
variables $\b$ and $V$, where, on shell, $\b$ is proportional to the Hubble parameter,
$\b =\gamma\frac{\dot{a}}{a}$ ($\gamma$ is the Barbero-Immirzi parameter) and $V$
is the physical volume of the cell ${\cal V}$. 
Note that $\b\in (-\infty ,\infty )$
and $V\ge 0$. The pair of conjugate variables satisfy $\{ \b ,V\} =4\pi G\gamma\, .$ The matter sector is given by
a scalar field $\phi$ and it momentum $p_\phi$, such that
$\{\phi ,p_{\phi}\}=1$.
 
This classical system is subject to the Hamiltonian constraint \cite{lqc}:
\be
 {\cal C}=N\left(-\frac{3}{8\pi G\gamma^2}V\b^2+\frac{{p_{\phi}}^2}{2V}\right)
\, .
\ee
If we choose $N=V$ the constraint takes simpler, classically equivalent form
\be
\tilde{\cal C}=-\frac{3}{8\pi G\gamma^2}V^2\b^2+\frac{p_{\phi}^2}{2} \label{con1}
\, .
\ee
The momentum $p_{\phi}$ is a Dirac observable and therefore, a constant of motion (we choose $p_{\phi}>0$ without loss of generality).
In order to construct one family of complete observables we should choose a suitable "clock" 
phase space function. From the form of the constraint $\tilde{\cal C}$ we can see that $\phi$ is a good
choice. Namely,
\be
\alpha^t_{\tilde{C}}(\phi )=\phi -p_{\phi}t\, , 
\ee
is an invertible function of $t$. For the other partial observable we choose $V$ and calculate its flow,
it turns out that it can be written in a closed form
\be
\alpha^t_{\tilde C}(V)=V\sum_{n=0}^{\infty}\frac{1}{n!}(-\frac{3}{\gamma}V\b t)^n=
Ve^{-\frac{3}{\gamma}V\b t}\, . 
\ee
Since from $\alpha^t_{\tilde{C}}(\phi )=\phi_0$ it follows that $t=\frac{\phi -\phi_0}{p_{\phi}}$,
and from (\ref{con1}) we obtain that $\frac{3}{\gamma}V|\b |\approx \kappa p_{\phi}$, where $\kappa =\sqrt{12\pi G}$, we can
express the corresponding one parameter family of complete observables as
\be
V\vert_{\phi_0}=Ve^{-\kappa ({\rm sgn}\b)(\phi -\phi_0)}\, .\label{obs1}
\ee

As our next example we shall consider a class of so called  `effective theories' characterized
by a parameter $\l$ with dimension of length, which can be thought of as a  `polymeric extension'
of the system (\ref{con1}).  This LQC
model that has received some of attention given that it can be solved both classical and quantum mechanically, 
for arbitrary values of $\l$. For details of the model see \cite{slqc}. 
The origin of $\l$ in the context of quantum theory is the discretization of $\hat{V}$, and
we take $\l$ as a free parameter and focus on the $\l$ dependence of the corresponding Dirac observables.
The effective theory on the scale $\l$ is defined by the constraint \cite{lqc,slqc} (we again take $N=V$)
\be
\tilde{\cal C}_{\l}=-\frac{3}{8\pi G\g^2}\frac{1}{\l^2}V^2\sin^2{(\l\b)}+\frac{{p_{\phi}}^2}{2}\label{con2}
\, ,
\ee
where $\b$ can be seen as being compactified, taking values in $\b\in (-\frac{\pi}{2\l},\frac{\pi}{2\l}]$.
This constraint implies that there is a minimum value of $V$ at every scale, $V_{b,\l}=\frac{\kappa}{3}p_{\phi}\g\l$
attained when $\b =\frac{\pi}{2\l}$. It turns out that all gauge trajectories have a bounce at
a point $(V,\b ,\phi ,p_{\phi})_{\l ,b} = (\frac{\kappa}{3}p_{\phi}\g\l ,\frac{\pi}{2\l}, \phi_{\l ,b} ,p_{\phi})$,
which depends on $\l$ and $p_{\phi}$. We note that $\phi_{\l ,b}$ is a finite number, that
behaves as $\ln{\l}$ as $\l\to 0$ \cite{CV}. This should be contrasted
with the behavior of the gauge trajectories obtained from (\ref{con1}), where there is no bounce and $V\to 0$ as
$\b\to \infty$. 

We note that $\lim_{\l\to 0}\tilde{\cal C}_{\l}=\tilde{\cal C}$. We are going to construct a set of complete
observables at every scale and analyze their behavior as $\l\to 0$, to be able to gain more insight into
the relation between the $k=0$ FRW cosmological model and the effective theories described above. As before,
we take $\phi$ and $V$ as two partial observables, and consider their flows. The flow of $\phi$ is
the one that we obtained previously, while the flow of $V$ takes the form 
\be
\alpha^t_{\tilde{C}_\l}(V)=V\left[ \sinh{(-\frac{3}{\gamma\l}Vt\sin{\l\b})}\cos{\l\b} +
\cosh{(\frac{3}{\gamma\l}Vt\sin{\l\b})}\right]\, ,
\ee
resulting in the following family of Dirac observables
\be
V_\l\vert_{\phi_0}=V_\l\bigl[\sinh{(-\kappa ({\rm sgn}\b)(\phi -\phi_0) })\,\cos{\l\b} +
\cosh{(\kappa ({\rm sgn}\b) (\phi -\phi_0))}\bigr]\, ,\label{obs2}
\ee
where the notation $V_\l$ indicates that we are at the scale $\l$. 

It can be shown easily \cite{CV}
that if we consider a gauge trajectory $V=V(\phi )$ in FRW model and compare it to $V=V_\l (\phi )$
it is always possible to choose some $\phi_0$ when the two trajectories are close to each other and
than see what happens during the evolution of the two systems. The two trajectories start close
to each other but start to diverge as one approaches the bounce of the effective theory. As
we make $\l$ smaller we can enlarge the interval of convergence, but eventually the two start
to diverge. We can also see this behavior at the level of Dirac observables. We see that
$V_\l\vert_{\phi_0}$ at bounce ($\b =\frac{\pi}{2\l}$) behaves as $V\cosh{(\kappa ({\rm sgn}\b) (\phi -\phi_0))}$
which should be compared to (\ref{obs1}). Of course it is just another point of view on the convergence
behavior that we already described at the level of gauge orbits.

Since the choice of a "clock" variable is not unique we can explore this freedom to search for the
set of complete observables that has better convergence properties than the one already constructed.
One possible "natural" candidate that represents a good parametrization of the gauge orbits is 
the cosmological proper time $\tau$. In this case we fix the lapse function as $N=1$ and one of the
corresponding equations of motion $\frac{d\tau}{d\phi}=\frac{V}{p_{\phi}}$ that, combined with the
rest of the equations can be solved leading to 
\be
\t = ({\rm sgn}\b )e^{\kappa ({\rm sgn}\b )\phi}\, ,
\ee
where the integration constant is fixed by the condition that $V(\t )\to 0$ as $\t\to 0$.
In this way we can construct $\t$ as a function on the phase space and choose it as the partial
observable that represents the "clock" function. We note that $\tau\ne 0$ on the constraint surface. 
The flow of $\tau$ is
\be
\alpha^t_{\tilde{C}}(\tau )=\tau e^{-\kappa ({\rm sgn}\b) p_{\phi}t}\, ,
\ee
and the one parameter family of complete observables is given by
\be
V\vert_{\t =\t_0}=V\frac{\t_0}{\t}\, .\label{obs3}
\ee
So, there are two sets of values that this observables can take, depending on the sign of $\t$. One of them, $\t >0$,
corresponds to the expanding universe, while the other one, for $\t <0$ corresponds to the universe in contraction.

We can use the same idea to construct new "clock", the $\l$-dependent partial observable in the effective theory 
\be 
\t_\l =\frac{\gamma\l}{3}\sinh{\left(\kappa ({\rm sgn}\b)\phi +\ln{\frac{3}{\gamma\l}}\right)}\, , 
\ee 
where the constant is fixed by the condition $\t_\l\to\t$ as $\l\to 0$. Note that at every scale the bounce occurs 
for the same value $\t_\l =0$. The flow of this function is 
\be 
\alpha^t_{\tilde{C}_\l}(\tau_\l )=\frac{\gamma\l}{3}\sinh{\left(\kappa ({\rm sgn}\b)(\phi -p_{\phi}t)+\ln{\frac{3}{\gamma\l}}\right)}\, . 
\ee
 When we calculate the family of corresponding complete observables we find that 
\be 
V_\l\vert_{\t_\l =\t_0}=\frac{1}{2}V\left[ (z-\frac{1}{z})\,\cos{\l\b}+(z+\frac{1}{z})\right]\, , \label{obs4} 
\ee 
with $z=(3\t_0+\sqrt{\gamma^2\l^2+9\t_0^2})(3\t_\l +\sqrt{\gamma^2\l^2+9\t_\l^2})^{-1}$. We can now compare the observables 
(\ref{obs4}) at the bounce, which now occurs for the same 
value of $\t_\l =0$. In the limit $\l\to 0$ they converge uniformly. At every scale $\l$ the gauge trajectory $V=V_\l (\t_\l)$ 
is a hyperbola, whose asymptotes are two gauge 
trajectories of a FRW model; two half-lines that correspond to the expanding ($p_{\phi}>0$) and contracting ($p_{\phi}<0$) 
branches \cite{CV}. The limit of the gauge trajectories 
when $\l\to 0$ exists, but is a non-differentiable curve. We can conclude that the new observables (\ref{obs3}) and (\ref{obs4}) 
are more suited for the description of the 
convergence of the effective theories. The limiting theory is a "singular bouncing" universe and {\it not} 
the classical GR dynamics, because $V_\l(0)\to 0$ as $\l\to 0$, and it does not belong to the phace space of the $k=0$
FRW cosmological model.

\section{Some results from the quantum theory}

The purpose of this section is to explore the relation between the Wheeler DeWitt and LQC when analyzing the dynamics of 
the volume as function of the internal time $\phi$.

Let us start with a very brief introduction of the quantum systems and their solutions, the details can be found in \cite{slqc}. The WDW theory of the quantum $k$=0 FRW model
coupled to a massless scalar field is obtained after the standard Schroedinger quantization of (\ref{con1}).
In the $(\b ,\phi )$ representation the Hamiltonian constraint takes the form
\be
\partial_\phi^2\underline{\Psi}(\b ,\phi )=12\pi G(\b\partial_\b )^2\underline{\Psi}(\b ,\phi )\, .
\ee
Physical states must satisfy this equation, and from the form of the equation it follows that the
scalar field can be used as a "clock" variable, just like in the classical theory. This constraint
can be rewritten as the Klein-Gordon equation if we replace $\b$ by  
$y\equiv\frac{1}{\kappa}\ln{\frac{\b}{\b_0}}$, with $\b_0$ arbitrary. 
Different choices of $\b_0$ yield unitarily equivalent quantum theories. We note that the change
$\b_0\to \b_0'=\a\b_0$ (where $\a$ is an arbitrary real number) corresponds to the change
$y\to y'=y-\frac{1}{\k}\ln{\a}$. The physical states take the form 
$\underline{\Psi}(y,\phi )=\underline{\Psi}_L(y_+)+\underline{\Psi}_R(y_-)$,
where $y_{\pm}=\phi\pm y$. We will show later that left (right) moving states, $\underline{\Psi}_L(y_+)$
($\underline{\Psi}_R(y_-)$) correspond to expanding (contracting) universes.

On the other hand, the solvable loop quantum cosmology (sLQC) model is obtained in the process of  
polymeric quantization of (\ref{con1}), where the discreteness parameter $\l$ is introduced in order to calculate curvature in terms of holonomies \cite{slqc} (Note that in this article the quantum theory was obtained by the standard loop quantization procedure without `polymerization'). One could also interpret  sLQC as being
obtained from the standard Schroedinger quantization of the effective constraint (\ref{con2}) (where the variable $\b$ is taken to be periodic).  
In the $(\b ,\phi )$ representation the
Hamiltonian constraint in sLQC takes the following form
\be
\partial_\phi^2\chi (\b ,\phi )=12\pi G\,\left(\frac{\sin{\l\b}}{\l}\partial_\b \right)^2\chi (\b ,\phi )\, .
\label{const_slqc}
\ee
We note that the physical states should be symmetric under the change of orientation of the physical co-triads.
As was shown in \cite{slqc}, for the WDW states this condition implies that we can restrict the analysis to the positive $\b$. For that
reason the domain of $\b$ in sLQC is chosen to be $(0,\frac{\pi}{\l})$. \footnote{In the classical $k=0$ FRW model we have
$\b\in (-\infty ,\infty )$ and for that reason we have chosen a symmetric domain for $\b$ in the classical effective theories, namely,
$\b\in (-\frac{\pi}{2\l}, \frac{\pi}{2\l}]$.}

The constraint (\ref{const_slqc}) can also be transformed to a Klein-Gordon equation if we introduce a new $\l$-dependent variable
$x=\frac{1}{\kappa}\ln{(\tan{\frac{\l \b}{2}})}$.
The physical states in sLQC are $\chi (x,\phi )=\chi_L (x_+)+\chi_R (x_-)$, where $x_{\pm}=\phi\pm x$.
The condition that the states should be symmetric under the change of the orientation of the physical co-triads 
leads to the restriction $\chi (-x,\phi )=-\chi (x,\phi )$. As a result, the
sLQC states are of the form
$
\chi (x,\phi )=\frac{1}{\sqrt{2}}(F(x_+)-F(x_-))\, ,
$
where $F$ is some function (for details see \cite{slqc}). 

In order to get the physical content of these theories we need to define the complete set of
physical observables. One possible choice in this case is $\hat{p}_{\phi}$ and $\hat{V}|_{\phi}$, in analogy with
classical theory. In the WDW theory these observables are defined in the following way
\be
\hat{p}_{\phi}\,\underline{\Psi}(y,\phi )=-i\partial_{\phi}\underline{\Psi}(y,\phi )= 
\sqrt{\underline\Theta}\;\underline{\Psi}(y,\phi )\, ,
\ee
\be
\hat{V}|_{\phi_0}\,\underline{\Psi}(y,\phi )=e^{i\sqrt{\underline\Theta}(\phi -\phi_0)}\,\hat{V}\;\underline{\Psi}(y,\phi_0)\, ,
\ee
where $\underline\Theta\equiv -\partial_y^2$. The same set of Dirac observables can be constructed in sLQC, their form can be
obtained from the above expressions after the substitution $\underline{\Psi}(y,\phi )\to \chi (x,\phi )$ and 
$\underline\Theta\to\Theta\equiv -\partial_x^2$.

Now, we want to compare the expectation values of $\hat{V}|_{\phi}$ in the appropriately chosen states in WDW theory and sLQC.
In WDW for left moving states one obtains \cite{slqc}
\be
(\underline{\Psi}_L,\hat{V}|_{\phi}\underline{\Psi}_L)_{{\rm phys}}=\tilde{V}_{0L}\,e^{\kappa\phi}\, ,\label{vol1}
\ee
where $\tilde{V}_{0L}$ is a constant depending on the initial conditions.
The expectation value tends to infinity as $\phi\to\infty$, for any left moving state, so that this sector corresponds 
to an expanding universe. Similarly, for the right moving solutions
\be
(\underline{\Psi}_R,\hat{V}|_{\phi}\underline{\Psi}_R)_{{\rm phys}}=\tilde{V}_{0R}\,e^{-\kappa\phi}\, ,\label{vol2}
\ee
and the right moving sector corresponds to a contracting universe. The explicit form of the constants is
\be
\tilde{V}_{0(L,R)}\equiv \frac{s}{\b_0}\int_{-\infty}^{\infty}{\d y\left| \frac{\d\underline{\Psi}_{(L,R)}(y)}{\d y}\right|^2
e^{\mp\k y}}\, ,\label{coef}
\ee
where $s=\frac{8\pi\g l_{\rm Pl}^2}{\k}$ .

In sLQC the expectation value of $\hat{V}|_\phi$ at the state $\chi (x,\phi )$ is of the form
\be
(\chi ,\hat{V}|_\phi\chi )_{\rm phy}= {\tilde V^{(\l )}}_+e^{\kappa\phi}+{\tilde V^{(\l )}}_-e^{-\kappa\phi}\, ,\label{vol3}
\ee
where ${\tilde V^{(\l )}}_{\pm}$ depend on the initial data and also on $\l$ 
\be
\tilde V^{(\l )}_{\pm}\equiv \frac{s\l}{2}\int_{-\infty}^{\infty}{\d x\left| \frac{\d F(x)}{\d x}\right|^2
e^{\mp\k x}}\, .\label{coef3}
\ee
The volume expectation value has a non-zero minimum value 
\be
V^{(\l )}_b=2\frac{\sqrt{\tilde V^{(\l )}_+\tilde V^{(\l )}_-}}{\Vert\chi_\l\Vert^2}\, ,\ \ \ {\rm for}\ \ \
\phi_b^{(\l)}=\frac{1}{2\k}\ln{\frac{\tilde V^{(\l )}_-}{{\tilde V^{(\l )}_+}}}\, ,
\ee
and it is symmetric about the bounce point $\langle\hat{V}|_\phi\rangle_{\phi_b+\phi} =V_b\cosh{\k\phi}$.

Let us first compare the predictions on two scales $\l$ and $\l '$. Let us suppose, without loss of generality, that $\l '<\l$. Since $x'\approx x-\frac{1}{\k}\ln{\frac{\l}{\l '}}$
for small $\b$ (that classically corresponds to the region where the universe is large), we can relate the states in two theories at different scales such that, for $\phi=0$, satisfy
the condition
\be
F_{\l '}(x')=F_\l (x'+\nu )\, ,
\ee
where $\nu =\frac{1}{\k}\ln{\frac{\l}{\l '}}>0$. These states have the same norm and the same expectation value for any power of
$p_{\phi}$. The relation between their volume expectation values can be obtained from
$$
{\tilde V}_\pm^{(\l ')}=\frac{s\l '}{2}\int_{-\infty}^{\infty}{\d x'\left| \frac{\d F_{\l '}(x')}{\d x'}\right|^2e^{\mp\k x'}}
=\frac{s\l '}{2}\int_{-\infty}^{\infty}{\d x'\left| \frac{\d F_{\l}(x'+\nu)}{\d x'}\right|^2e^{\mp\k x'}}=\frac{\l '}{\l}e^{\pm\k\nu}{\tilde V}_\pm^{(\l )}
$$
so that
\be
\tilde V^{(\l ')}_+=\tilde V^{(\l )}_+\ \ \ {\rm and}\ \ \  
\tilde V^{(\l ')}_-=\left( \frac{\l '}{\l}\right)^2\ \tilde V^{(\l )}_-\, .
\ee
Thus, with this prescription we are `synchronizing' the states in such a way that the expanding branch coincides for all values of the parameter $\l$. Clearly one should expect that these states approximate a WDW state for large positive values of $\phi$. Note also that, for $\phi =0$,
$$
\langle\hat{V}|_0\rangle^{(\l )}-\langle\hat{V}|_0\rangle^{(\l ')}=\left[ 1-\left( \frac{\l '}{\l}\right)^2\right] \tilde V_-^{(\l )}\, ,
$$
and the volume expectation values on two scales are close to each other for large $\phi$ but they grow apart as
$\phi\to\phi_b$ (and even further for $\phi<\phi_b$). 

The minimum values of volume are related as
\be
V_b^{(\l ')}=\frac{\l '}{\l}V_b^{(\l )}\, ,\ \ \ {\rm where}\ \ \ \phi_b^{(\l ')}=\phi_b^{(\l )}-\nu\, .
\ee
We see that $\phi_b^{(\l ')}\to -\infty$ as $\l '\to 0$. As we have synchronized the states in such a way that the dynamics coincides for 
large positive values of the scalar field, what we are doing is to `push' the bounce to more negative values of the internal time as we 
decrease the value of $\l '$. Let us now see how we can `synchronize' states during the contracting phase, in such a way that now the bounce 
gets `pushed' to large values of the internal time $\phi$.

For that, let us choose the states such that, for $\phi =0$,
\be
F_{\l '}(x')=F_\l (x'-\nu )\, ,
\ee
we obtain, following the same steps as above, that
\be
\tilde V^{(\l ')}_+=\left( \frac{\l '}{\l}\right)^2\ \tilde V^{(\l )}_+\ \ \ {\rm and}\ \ \  
\tilde V^{(\l ')}_-=\tilde V^{(\l )}_-\, ,
\ee
and we see that now the expectation values are close for large negative $\phi$ and they grow apart as $\phi$ approaches the corresponding
bounce point. In this case
\be
\phi_b^{(\l ')}=\phi_b^{(\l )}+\nu\, ,
\ee
and $\phi_b^{(\l ')}\to\infty$ as $\l '\to 0$,
which is what we wanted to get. Let us now compare sLQC states with those of the Wheeler DeWitt theory.

In order to be able to compare predictions from WDW theory and sLQC we should choose the states in both theories that
are `close' to each other for small $\b$, in the sense that the expectation
values of $\hat{V}|_\phi$ in these states should be close to each other. As we shall see, one can define such prescription 
for both branches of the WDW theory.
First we should initially (for $\phi =0$) choose a state $\underline{\Psi}_L (y)$ such that 
$ {\tilde V}_-$ is small, when calculated on this state. In order to construct the corresponding state in sLQC we first note that 
\be
x\approx\frac{1}{\k}
\ln{\frac{\l\b}{2}}=y-\frac{1}{\k}\ln{\frac{2}{\l\b_0}}\, ,\ \ \ \rm{when}\ \ \b\to 0\, . 
\ee
As a result, if we consider states such that
\be
F_{\l}(x)\equiv \underline{\Psi}_L (x+\mu )\, ,
\ee
where $\mu =\frac{1}{\k}\ln{\frac{2}{\l\b_0}}$,
then the volume expectation values in these states
are close. That is, for $\phi =0$, they can be arbitrarily close for an appropriate choice of $\l$. 
We also note that these states have the same norm and that the corresponding expectation values
of $\hat{p}_{\phi}$ are the same \cite{slqc}. Using this result we can calculate the explicit $\l$ dependence of
${\tilde V^{(\l )}}_{\pm}$,
\be
\tilde V^{(\l )}_{\pm}= \frac{s\l}{2}e^{\pm\k\mu}\int_{-\infty}^{\infty}{\d x
\left| \frac{\d\underline{\Psi}_L (x)}{\d x}\right|^2 e^{\mp\k x}}\equiv \l e^{\pm\k\mu}K_{\pm}\, ,
\ee
(notice that $K_{\pm}$ do not depend on $\l$), so that
\be
{\tilde V^{(\l )}}_{+}=\tilde{V}_{0L}\, , \ \ \ {\tilde V^{(\l )}}_{-}=\l^2 \frac{\b_0^2}{4}\,\tilde{V}_{0R}
\ee
we can now see that ${\tilde V^{(\l )}}_{-}\to 0$ as $\l\to 0$, so that for any $\phi$ fixed, the dominant part
of (\ref{vol3}) is the same as (\ref{vol1}) in this limit. 
 
Similarly, if we initially choose a state $\underline{\Psi}_R (y)$ such that $ {\tilde V}_+$ is small,
than the condition
\be
F_{\l}(x)\equiv \underline{\Psi}_R (x-\mu )\, ,
\ee
leads to
\be
{\tilde V^{(\l )}}_{+}=\l^2\frac{\b_0^2}{4}\tilde{V}_{0L}\, , \ \ \ {\tilde V^{(\l )}}_{-}=\tilde{V}_{0R}\, ,
\ee
so, in this case ${\tilde V^{(\l )}}_{+}\to 0$ as $\l\to 0$. We can see that there is a convergence of the
coefficients of (\ref{coef3}) as $\l\to 0$ to their WDW counterparts (\ref{coef}), but the
convergence of the expectation values (\ref{vol3}) in non-uniform, due to a divergent behavior in the
limit $\phi\to\pm\infty$. 

Motivated by the results of the classical analysis presented in the previous section, we could try to construct
another set of complete observables in quantum theory that would have better convergence properties in the limit $\l\to 0$.
This could be $\hat{V}|_{\t}$. In order to construct these observables we should first introduce the proper
time operator on the corresponding Hilbert space, and then construct physical operators corresponding to, say, 
volume at a given proper time. This is a non trivial task that we leave for future work.

\section{Discussion} \label{sec:5} 

The problem of taking the continuum limit of constrained polymeric quantum theories and loop quantum cosmology in 
particular is an open and interesting question. We have available some partial answers, that we have revisited in 
this contribution. On the one hand, a classical analysis of the related ``effective theory'' clearly shows that the 
issue of convergence depends very strongly on the choice of time. Only with respect to a particular $\l$-dependent 
internal time can the dynamics have a chance to converge. For the system under consideration we have shown that the 
cosmic proper time is a valid choice with respect to which the limit can be studied. As we saw, even when the 
limit exists and is well defined, the limiting dynamics is non-differentiable. That is, the resulting limit does not correspond to a 
cosmology satisfying reasonable equations of motion but rather suffers from a "singular bounce". 

On the quantum front, the result that is available pertains to the evolution of the expectation value of volume with respect 
to the scalar field $\phi$ that can be used as an internal time. As we showed in detail, from the loop quantum theory one 
can indeed recover both the contracting and expanding branches of the Wheeler DeWitt theory, but strictly 
speaking, there is no convergence between the theories since the limit is not uniform. One should note that this result in 
the quantum domain is fully consistent with the classical analysis in which it was shown that using the scalar 
field as the internal time does not lead to convergence of the limit $\l\to 0$ for the effective theories.

The step that is clearly missing in order to have a converging quantum theory is to translate the results of the classical 
effective theory to the quantum theory. That is, one would 
like to have a consistent procedure to define the quantum dynamics in terms of proper time. The difficulty with this is 
that proper time is itself a rather non-trivial function on 
phase space, for which one needs to construct a well defined operator (See, for instance, \cite{Paw} for a proposal). 
Furthermore, one needs to define a relational dynamics for the volume in terms of the proper time operator. One possibility 
would be to define the relational dynamics in terms of expectation values. For instance, for the volume operator that one 
uses to explore singularity resolution in LQC, one could consider the evolution of the expectation value of the volume 
operator at a given value of the expectation value of the {\it proper time} operator. One would expect that quantity to be 
well defined in the limit, for the model we have here studied.

We shall leave a detailed analysis of the problem of defining the continuum limit for quantum constrained systems and the 
analysis of these simple systems for a future publication.

\section*{Acknowledgments}
\noindent
We thank J.A. Zapata for discussions and comments during the early part of this project. 
This work was in part supported by DGAPA-UNAM IN103610 grant, by NSF
PHY0854743, the Eberly Research Funds of Penn State and by CIC, UMSNH.

\end{document}